# The Number Theoretic Hilbert Transform

Subhash Kak[1]

**Abstract**
This paper presents a general expression for a number-theoretic Hilbert transform (NHT). The transformations preserve the circulant nature of the discrete Hilbert transform (DHT) matrix together with alternating values in each row being zero and non-zero. Specific examples for 4-point, 6-point, and 8-point NHT are provided. The NHT transformation can be used as a primitive to create cryptographically useful scrambling transformations.

## Introduction

The Hilbert transform has important applications in signal theory where it is used along with the Fourier transform. Its discrete version, the discrete Hilbert transform (DHT) [1], is used in communication theory, modulation, determination of instantaneous frequency, signal processing [2],[3] and cryptography [4]. The DHT formulas were developed in different ways for data of finite length [5],[6], by taking the infinite data sequence to be periodic. But as of now number theoretic version of the DHT are not known and such a version could be developed along the same lines as number theoretic Fourier transform. The discrete Fourier transform is defined over a ring whereas its number theoretic version is defined in a finite field and the computations for the number theoretic transform are with respect to a suitable modulus.

One motivation to seek formulas for NHT is its potential use in scrambling [7]-[10]. A series of NHTs, together with suitable number theoretic transforms, can be used on the data to create a hard to invert transformation. The small-block NHT formulas can be used in a braided form [11]. Such a scheme can be used as a primitive in scrambling and hashing transformations. In classical applications it can be used directly; in quantum cryptography, it can be used to choose random bases [12], or random polarization angles [13],[14].

Here we present a general expression for the number theoretic Hilbert transform (NHT) that has a form similar to that of DHT. The exact forms of 4-point, 6-point, and 8-point NHT have been derived by reasoning similar to what leads to the number theoretic discrete Fourier transform [15]-[17]. The NHT formulas have an exact inverse unlike the approximate inverse proposed earlier [18]. Two versions of NHT are presented: (i) in which the inverse transform is the transpose of the direct transform, and (ii) in which the inverse transform is identical to the direct transform. The second form does not always exist.

[1] Department of Computer Science, Oklahoma State University, Stillwater, OK 74078

## The Basic Discrete Hilbert Transform

The formulas for the basic Discrete Hilbert Transform (DHT) of discrete data *f(n)* where *n = (-∞,...,-1,0,1,...,∞)* are as follows [1]:

$$DHT\{f(n)\} = g(k) = \begin{cases} \dfrac{2}{\pi} \sum_{n \text{ odd}} \dfrac{f(n)}{k-n}; & k \text{ even} \\ \dfrac{2}{\pi} \sum_{n \text{ even}} \dfrac{f(n)}{k-n}; & k \text{ odd} \end{cases} \qquad (1)$$

The inverse Discrete Hilbert Transform (DHT) is given as:

$$f(n) = \begin{cases} -\dfrac{2}{\pi} \sum_{k \text{ odd}} \dfrac{g(k)}{n-k}; & n \text{ even} \\ -\dfrac{2}{\pi} \sum_{k \text{ even}} \dfrac{g(k)}{n-k}; & n \text{ odd} \end{cases} \qquad (2)$$

The DHT matrix is a circulant matrix. The values given in the first row (in which non-zero and zero values alternate) are shifted by one as one goes to lower rows. The transformation for the DHT matrix H, input vector F, and the output vector G, can be written as

$$G = HF \qquad (3)$$

For example, we can write:

$$G = \dfrac{2}{\pi} \begin{bmatrix} 0 & \dfrac{1}{-1} & 0 & \dfrac{1}{-3} & 0 & \dfrac{1}{-5} & 0 & \dfrac{1}{-7} & \cdot \\ \dfrac{1}{1} & 0 & \dfrac{1}{-1} & 0 & \dfrac{1}{-3} & 0 & \cdot & \cdot & \cdot \\ 0 & \dfrac{1}{1} & 0 & \dfrac{1}{-1} & 0 & \dfrac{1}{-3} & \cdot & \cdot & \cdot \\ \dfrac{1}{3} & 0 & \dfrac{1}{1} & 0 & \dfrac{1}{-1} & 0 & \cdot & \cdot & \cdot \\ 0 & \dfrac{1}{3} & 0 & \dfrac{1}{1} & 0 & \dfrac{1}{-1} & \cdot & \cdot & \cdot \\ \dfrac{1}{5} & 0 & \dfrac{1}{3} & 0 & \dfrac{1}{1} & 0 & \cdot & \cdot & \cdot \\ 0 & \cdot & \cdot & \cdot & \cdot & \cdot & \cdot & \cdot & \cdot \\ \dfrac{1}{7} & \cdot & \cdot & \cdot & \cdot & \cdot & \cdot & \cdot & \cdot \\ \cdot & \cdot & \cdot & \cdot & \cdot & \cdot & \cdot & \cdot & \cdot \end{bmatrix} F$$

It is easy to convert this into a corresponding number theoretic forward transformation if the computations are done modulo a large enough prime that is a bit larger than the matrix number of rows, but the problem is that the inverse transformation must consider an infinite sized matrix. By restricting the size of the inverse matrix to be the same size as the forward matrix, we are introducing an approximation. We cannot use finite versions of DHT as starting point for number theoretic NHT because the values within the matrix are irrational [6].

Therefore, the development of a number theoretic transformation must be considered somewhat differently.

## Number Theoretic DHT (NHT)

We propose that instead of using the exact form of the DHT matrix and converting it into corresponding integers for numbers modulo an appropriate prime, we should merely consider the form of the matrix with its alternate zeros and non-zeros. Incidentally, this structure was the motivation to seek multilayered and unusual arrays for matrix multiplication [19],[20].

Consider the data block to be *F* and the NHT transform of it to be *G*. We propose the following matrix N as the general form of NHT:

$$G = NF \mod m \tag{4}$$

where

$$N = \begin{bmatrix} 0 & a & 0 & b & 0 & c & 0 & . & . & k \\ k & 0 & a & 0 & b & 0 & c & 0 & . & . \\ . & k & 0 & a & 0 & b & 0 & c & 0 & . \\ . & . & k & 0 & a & 0 & b & 0 & c & 0 \\ 0 & . & . & k & 0 & a & 0 & b & 0 & c \\ c & 0 & . & . & k & 0 & a & 0 & b & 0 \\ 0 & c & 0 & . & . & k & 0 & a & 0 & b \\ b & 0 & c & 0 & . & . & k & 0 & a & 0 \\ 0 & b & 0 & c & 0 & . & . & k & 0 & a \\ a & 0 & b & 0 & c & 0 & . & . & k & 0 \end{bmatrix} \mod m \tag{5}$$

and *m* is an appropriate value of the modulus. How the modulus is to be found will be explained when specific cases of NHT are taken. The inverse of the transform is

$$F = N^T G \mod m \tag{6}$$

We now describe 4-point, 6-point, and 8-point versions of the NHT matrix.

## 4-point NHT

Let

$$N = \begin{bmatrix} 0 & a & 0 & b \\ b & 0 & a & 0 \\ 0 & b & 0 & a \\ a & 0 & b & 0 \end{bmatrix} \qquad (7)$$

By a straightforward calculation, we obtain:

**Theorem 1.** $NN^T = (a^2 + b^2)I \mod 2ab$ (8)

As example, consider $a=3/2$ and $b=5$. This gives us m=15 and $NN^T=(9/4+25)$ mod 15. This can be simplified to $NN^T$=61 mod 15, which is equivalent to $NN^T$=1 mod 15.

**Theorem 2.** $NN = 2ab \mod(a^2 + b^2)$ (9)

This provides a dual transformation matrix.

## 6-point NHT

Let a particular form of the 6-point matrix for DHT be given by the following matrix:

$$N = \begin{bmatrix} 0 & a & 0 & a+2 & 0 & a+4 \\ a+4 & 0 & a & 0 & a+2 & 0 \\ 0 & a+4 & 0 & a & 0 & a+2 \\ a+2 & 0 & a+4 & 0 & a & 0 \\ 0 & a+2 & 0 & a+4 & 0 & a \\ a & 0 & a+2 & 0 & a+4 & 0 \end{bmatrix} \qquad (10)$$

It is assumed that $a, a+2, a+4$ are relatively prime. Note that in addition to the alternating 0s, the increment in the values of *a* is by 2. The denominator in the basic Hilbert transform also changes by 2. The reason why we are doing so in the numerator rather than the denominator is for convenience. In any event, we will also consider another form of NHT where the numbers are arbitrary.

**Theorem 3.** $NN^T = 12I \mod 3a^2 + 12a + 8$. (11)

*Proof.* The proof of this follows from the fact that the diagonal terms of the matrix $NN^T$ are $(3a^2 + 12a + 20)$ and the off-diagonal terms of the circulant matrix are $(3a^2 + 12a + 8)$. To normalize the transformation matrix, each term of (11) will be divided by $\sqrt{12}$. ∎

**Theorem 4.** $NN = -12I \mod 3a^2 + 12a + 16$. (12)

*Proof.* The proof of this is similar to that of the previous theorem.∎

Theorems 3 and 4 provide dual transformations that are different in the modulus chosen. The two transformations will be distinguished by the use of subscript. Thus $N_1$ is the transformation where the inverse is the transpose and $N_2$ is the transformation where the inverse is the same as the forward transformation. Once the matrices have been normalized, we can write:

$$G = N_1 F \mod 3a^2 + 12a + 8$$
$$F = N_1^T G \mod 3a^2 + 12a + 8$$
(13)

For the second type:

$$G = N_2 F \mod 3a^2 + 12a + 16$$
$$F = N_2 G \mod 3a^2 + 12a + 16$$
(14)

By choosing different values of *a*, different transforms will be obtained. Some examples of this follow.

*Example 1.* Consider the elements 0, 1, 0, 2, 0, 3 in the first row of the matrix (10). Using the results of the two theorems, the moduli for computation are 23 for transpose inverse and 31 for the same inverse as the forward transformation. To normalize the matrices, the elements will have to be divided by $\sqrt{12}$ and $\sqrt{-12}$, respectively. Since $\sqrt{12} \mod 23$ is 9 and $\sqrt{-12} \mod 31$ is also 9, the circulant matrices obtained will be 0, 1/9, 0, 2/9, 0, 3/9 mod 23 and 0, 1/9, 0, 2/9, 0, 3/9 mod 31, respectively.

This yields the first row to be 0, 18, 0, 13, 0, 8 for the transformation where the inverse is the transpose of the forward transformation.

$$\begin{bmatrix} g(0) \\ g(1) \\ g(2) \\ g(3) \\ g(4) \\ g(5) \end{bmatrix} = \begin{bmatrix} 0 & 18 & 0 & 13 & 0 & 8 \\ 8 & 0 & 18 & 0 & 13 & 0 \\ 0 & 8 & 0 & 18 & 0 & 13 \\ 13 & 0 & 8 & 0 & 18 & 0 \\ 0 & 13 & 0 & 8 & 0 & 18 \\ 18 & 0 & 13 & 0 & 8 & 0 \end{bmatrix} \begin{bmatrix} f(0) \\ f(1) \\ f(2) \\ f(3) \\ f(4) \\ f(5) \end{bmatrix} \mod 23 \quad (15)$$

*Example 2.* Now we consider the dual transformation when the inverse matrix is identical to the forward matrix:

$$\begin{bmatrix} g(0) \\ g(1) \\ g(2) \\ g(3) \\ g(4) \\ g(5) \end{bmatrix} = \frac{1}{9} \begin{bmatrix} 0 & 1 & 0 & 3 & 0 & 5 \\ 5 & 0 & 1 & 0 & 3 & 0 \\ 0 & 5 & 0 & 1 & 0 & 3 \\ 3 & 0 & 5 & 0 & 1 & 0 \\ 0 & 3 & 0 & 5 & 0 & 1 \\ 1 & 0 & 3 & 0 & 5 & 0 \end{bmatrix} \begin{bmatrix} f(0) \\ f(1) \\ f(2) \\ f(3) \\ f(4) \\ f(5) \end{bmatrix} \mod 31 \qquad (16)$$

The transformation matrix may be written down equivalently as:

$$\begin{bmatrix} 0 & 7 & 0 & 21 & 0 & 4 \\ 4 & 0 & 7 & 0 & 21 & 0 \\ 0 & 4 & 0 & 7 & 0 & 21 \\ 21 & 0 & 4 & 0 & 7 & 0 \\ 0 & 21 & 0 & 4 & 0 & 7 \\ 7 & 0 & 21 & 0 & 4 & 0 \end{bmatrix} \mod 31 \qquad (17)$$

This may be used in conjunction with the number theoretic (Fourier) transformation:

$$L = \begin{bmatrix} 1 & 1 & 1 & 1 & 1 & 1 \\ 1 & 6 & 5 & 30 & 25 & 26 \\ 1 & 5 & 25 & 1 & 5 & 25 \\ 1 & 30 & 1 & 30 & 1 & 30 \\ 1 & 25 & 5 & 1 & 25 & 5 \\ 1 & 26 & 25 & 30 & 5 & 6 \end{bmatrix} \mod 31 \qquad (18)$$

which is from the fact that the order of 6 mod 31 is 6. The entries in row $i$ (counted as 0, 1, 2,..) are *1, $6^i$, $6^{2i}$*, and so on. The entries of $L^{-1}$, $i$th row, are *1, $6^{-i}$, $6^{-2i}$*, …. The inverse matrix is thus:

$$L^{-1} = \frac{1}{6} \begin{bmatrix} 1 & 1 & 1 & 1 & 1 & 1 \\ 1 & 26 & 25 & 30 & 5 & 6 \\ 1 & 25 & 5 & 1 & 25 & 5 \\ 1 & 30 & 1 & 30 & 1 & 30 \\ 1 & 5 & 25 & 1 & 5 & 25 \\ 1 & 6 & 5 & 30 & 25 & 26 \end{bmatrix} \mod 31 \qquad (19)$$

*Example 3*. Consider 0, 2, 0, 4, 0, 6 in the first row. Since the three numbers are not mutually prime, one can divide the moduli for the two cases that turn out to be 44 and 52 by the square of the gcd of the numbers. This gives us the moduli as 11 and 13 for the transforms $N_1$ and $N_2$, respectively. Below, we provide further details on the $N_2$ transformation.

$$\begin{bmatrix} g(0) \\ g(1) \\ g(2) \\ g(3) \\ g(4) \\ g(5) \end{bmatrix} = \begin{bmatrix} 0 & 2 & 0 & 4 & 0 & 6 \\ 6 & 0 & 2 & 0 & 4 & 0 \\ 0 & 6 & 0 & 2 & 0 & 4 \\ 4 & 0 & 6 & 0 & 2 & 0 \\ 0 & 4 & 0 & 6 & 0 & 2 \\ 2 & 0 & 4 & 0 & 6 & 0 \end{bmatrix} \begin{bmatrix} f(0) \\ f(1) \\ f(2) \\ f(3) \\ f(4) \\ f(5) \end{bmatrix} \mod 13 \qquad (20)$$

Some examples of NHT of equation 20 are given in the table below:

Table 1. Some NHT pairs for mapping (20)

|   | f(n) | g(n) |
|---|---|---|
| 1 | 1, 1, 1, 1, 1, 1 | 12, 12, 12, 12, 12, 12 |
| 2 | 1, 1, 1, 0, 0, 0 | 2, 8, 6, 10, 4, 6 |
| 3 | 0, 1, 1, 1, 0, 0 | 6, 2, 8, 6, 10, 4 |
| 4 | 0, 0, 1, 1, 1, 0 | 4, 6, 2, 8, 6, 10 |
| 5 | 1, 1, 0, 0, 1, 1 | 8, 10, 10, 6, 6, 8 |
| 6 | 0, 0, 1, 1, 0, 0 | 4, 2, 2, 6, 6, 4 |
| 7 | 1, 2, 3, 4, 5, 6 | 4, 6, 5, 6, 5, 5 |
| 8 | 3, 4, 5, 6, 1, 2 | 5, 6, 5, 5, 4, 6 |
| 9 | 2, 3, 4, 5, 6, 7 | 3, 5, 4, 5, 4, 4 |

By considering examples 2 and 3 in the Table above, we have the result that a right circular shift of the data results in a corresponding left circular shift of the transform. This may be written down as follows:

If *DHT[f(k)] = g(k)*, then *DHT[f(k+i)] = g(k-i)*

Below we provide some additional pairs of blocks that are mutual transforms.

|   | f(n) | g(n) |
|---|---|---|
| 11 | 1, 0, 0, 0, 0, 0 | 0, 6, 0, 4, 0, 2 |
| 12 | 0, 6, 0, 4, 0, 2 | 1, 0, 0, 0, 0, 0 |
| 13 | 1, 2, 3, 3, 2, 1 | 9, 7, 9, 0, 2, 0 |
| 14 | 3, 2, 1, 1, 2, 3 | 0, 2, 0, 9, 7, 9 |
| 15 | 4, 4, 4, 4, 4, 4 | 9, 9, 9, 9, 9, 9 |

Since the transform of [1, 0, 0, 0, 0, 0] is [0, 6, 0, 4, 0, 2], one can use result this together with the shift property and linearity to determine the transform of any sequence.

## General Form of NHT

We assume that the numbers in the first row are $0, a, 0, a+k, 0, a+l$. We obtain the following result by a straightforward analysis.

**Theorem 5.** $NN^T = (k^2 + l^2 - kl)I \mod 3a^2 + 2ak + 2al + kl$. (21)

Example 4. Let $a=1$, $k=4$, and $l=6$. By substituting in (21), we get the modulus of 47 and the above equation is $NN^T = 28I \mod 47$. Since $\sqrt{28} \mod 47 = 13$, we can write:

$$\begin{bmatrix} g(0) \\ g(1) \\ g(2) \\ g(3) \\ g(4) \\ g(5) \end{bmatrix} = \frac{1}{13} \begin{bmatrix} 0 & 1 & 0 & 5 & 0 & 7 \\ 7 & 0 & 1 & 0 & 5 & 0 \\ 0 & 7 & 0 & 1 & 0 & 5 \\ 5 & 0 & 7 & 0 & 1 & 0 \\ 0 & 5 & 0 & 7 & 0 & 1 \\ 1 & 0 & 5 & 0 & 7 & 0 \end{bmatrix} \begin{bmatrix} f(0) \\ f(1) \\ f(2) \\ f(3) \\ f(4) \\ f(5) \end{bmatrix} \mod 47 \qquad (22)$$

The inverse equation is the transpose of this forward equation and it also carries the division by 13. The transformation matrix may be written down equivalently as:

$$\begin{bmatrix} 0 & 18 & 0 & 43 & 0 & 32 \\ 32 & 0 & 18 & 0 & 43 & 0 \\ 0 & 32 & 0 & 18 & 0 & 43 \\ 43 & 0 & 32 & 0 & 18 & 0 \\ 0 & 43 & 0 & 32 & 0 & 18 \\ 18 & 0 & 43 & 0 & 32 & 0 \end{bmatrix} \mod 47 \qquad (23)$$

## 8-Point NHT

Given the first row of the 8-point NHT matrix is 0, a, 0, b, 0, c, 0, d, a straightforward analysis shows that for an inverse to exist,

$$ab + bc + cd + da = 2ac + 2bd \qquad (24)$$

The diagonal will then have the term $a^2 + b^2 + c^2 + d^2$ and the modulus can be taken to be either side of equation (24).

This equation can be satisfied by infinity of solutions. Suppose, we arbitrarily choose $a=1, b=-1, c=3$, then a simple calculation shows that $d=5/3$. We can thus write the 8-point forward NHT transformation as $G = \frac{1}{2} NF \mod 24$ as shown below:

$$\begin{bmatrix} g(0) \\ g(1) \\ g(2) \\ g(3) \\ g(4) \\ g(5) \\ g(6) \\ g(7) \end{bmatrix} = \frac{1}{2} \begin{bmatrix} 0 & 3 & 0 & -3 & 0 & 9 & 0 & 5 \\ 5 & 0 & 3 & 0 & -3 & 0 & 9 & 0 \\ 0 & 5 & 0 & 3 & 0 & -3 & 0 & 9 \\ 9 & 0 & 5 & 0 & 3 & 0 & -3 & 0 \\ 0 & 9 & 0 & 5 & 0 & 3 & 0 & -3 \\ -3 & 0 & 9 & 0 & 5 & 0 & 3 & 0 \\ 0 & -3 & 0 & 9 & 0 & 5 & 0 & 3 \\ 3 & 0 & -3 & 0 & 9 & 0 & 5 & 0 \end{bmatrix} \begin{bmatrix} f(0) \\ f(1) \\ f(2) \\ f(3) \\ f(4) \\ f(5) \\ f(6) \\ f(7) \end{bmatrix} \mod 24 \quad (25)$$

It is easy to check that $NN^T = 4I \mod 24$. Even though the modulus is a not a prime number, the inverse is defined because all the entries in the matrix are relatively prime to the modulus. Table 2 provides data blocks and the corresponding NHT blocks.

**Table 2**. 8-point NHT for (24)

|   | f(n) | g(n) |
|---|------|------|
| 1 | 1, 1, 1, 1, 1, 1, 1, 1 | 7, 7, 7, 7, 7, 7, 7, 7 |
| 2 | 1, 1, 1, 0, 0, 0, 0, 0 | 3/2,4,5/2,7,9/2,3,-3/2,0 |
| 3 | 0, 1, 1, 1, 0, 0, 0, 0 | 0, 3/2,4,5/2,7,9/2,3,-3/2 |
| 4 | 0, 0, 1, 1, 1, 1, 0, 0 | 3, 0, 0, 4, 4, 7, 7, 3 |
| 5 | 1, 1, 0, 0, 0, 0, 1, 1 | 4, 7, 7, 3, 3, 0, 0, 4 |
| 6 | 0, 0, 1, 1, 0, 0, 1, 1 | 1, 6, 6, 1, 1, 6, 6, 1 |
| 7 | 1, 2, 3, 4, 5, 6, 7, 8 | 8, 10, 2, 9, 4, 11, 6, 1 |
| 8 | 1, 2, 3, 4, 4, 3, 2, 1 | 13, 10, 11, 15, 22, 1, 0, 20 |

Some other choices for the parameters of the 8-point NHT are given below.

$a = 5, b = -5, c = 10, d = 7$ and the modulus is 30

$a = 10, b = -5, c = 5, d = 7$ and the modulus is 30

These two, of course, are variants of the same matrix. The formula in this case is:

$$NN^T = 19I \mod 30$$

Large values of the moduli may also be obtained as in the examples below:

A) $a = 77, b = -7, c = 7, d = 17$ and the modulus is 840 and the identity is:

$$NN^T = 436I \mod 840$$

B) $a = 36, b = -9, c = 27, d = 31$ and the modulus is 1386 and the identity is:

$$NN^T = 295I \mod 1386$$

## Conclusions

We have presented a general expression for a number-theoretic discrete Hilbert transform (NHT). The transformations preserve the circulant nature of the discrete Hilbert transform (DHT) matrix together with alternating values in each row being zero and non-zero. Specific examples for 4-point, 6-point, and 8-point NHT are provided. The NHT transformation can be used as a primitive to create cryptographically useful scrambling transformations for use in classical and quantum cryptography.

Apart from the determination of NHT for higher-sized blocks, one needs to explore the specific forms of NHT that can be used easily in conjunction with NTTs.

*Acknowledgement.* This research was supported in part by the National Science Foundation grant CNS-1117068.